# A limit on exotic $\nu_\mu$ - $\nu_\tau$ oscillations derived from results of the OPERA experiment

P.F.Loverre
*Università di Roma "La Sapienza" and INFN Sezione di Roma, Roma, Italy*

*Abstract*

The OPERA experiment searching for $\nu_\mu - \nu_\tau$ oscillations in the CERN to Gran Sasso $\nu_\mu$ beam, has detected so far three candidates of $\nu_\tau$ charged current interactions. All three events are attributed by the authors to the standard $\nu_\mu - \nu_\tau$ oscillation at the atmospheric $\Delta m^2$. It is then possible to constrain the number of additional interactions, which could e.g. result from non-standard $\nu_\mu - \nu_\tau$ oscillations involving more than 3 neutrino mass states. In this paper, a 90% CL upper limit of 5.5 events is set on the number of additional $\nu_\tau$ interactions. Using for the *exotic* oscillation a probability of the form $P=\sin^2(2\theta_X)\sin^2(1.27\Delta m^2_X(L/E))$, the u.l. of 5.5 events is converted into a limit on $\sin^2(2\theta_X)$. For $\Delta m^2_X$ larger than 0.02 eV$^2$, the constraint is $\sin^2(2\theta_X) < 8.2\times10^{-2}$ .

________________________________________________

The OPERA experiment [1], operating in a pure $\nu_\mu$ beam produced at CERN and directed at the detector at the LNGS laboratory, 732 km away, has recently reported [2] the observation of three $\nu_\tau$ charged-current (CC) interactions, identified through the direct detection of the τ lepton decay. In the OPERA paper, the observation of three events is shown to be consistent with the expectation for the $\nu_\mu - \nu_\tau$ oscillation, as computed in a two-flavour oscillation scheme with the parameters of the atmospheric sector. It follows, that the OPERA experiment does not observe any excess of the number of $\nu_\tau$ interactions, with respect to what expected from the standard model with three neutrino flavours and three neutrino mass states. Here, the OPERA result is then used to compute an upper limit to the rate of *additional* $\nu_\tau$ interactions. Additional $\nu_\tau$ interactions could be produced for instance, by $\nu_\mu - \nu_\tau$ oscillations regulated by a $\Delta m^2$ larger than the atmospheric one, and related to the existence of one or more sterile neutrinos, as discussed in various exotic models (see e.g. ref.[3] and references therein). It is worth noting that OPERA, because of the high threshold for the production of the τ lepton, uses a beam of relatively high energy, $<E_\nu> = 17$ GeV. At that energy, and with L=732 km, the oscillation probability term $\sin^2(1.27\Delta m^2(L/E))$ is very small, when computed at the atmospheric $\Delta m^2$ ($\Delta m^2_{atm}=2.32\times10^{-3}$ eV$^2$). One has: $\sin^2(1.27\times2.32\times10^{-3}\times(732/17))$ = 0.016. The same term, for $\Delta m^2$ larger than 0.10 eV$^2$, would be averaged over the beam to the value 0.5, and would then be larger by a factor 30. It is then clear that the sensitivity of OPERA to oscillations regulated by a $\sin^2(1.27\Delta m^2(L/E))$ term, would be much greater for $\Delta m^2$ values larger than the atmospheric one, and that correspondingly, the absence of an excess of $\nu_\tau$ interactions allows to set limits on such oscillations.

The calculation of the limit goes through the following steps. The OPERA paper [2] reports the observation of 3 $\nu_\tau$ interactions, in a sample where 1.88 events are expected. Of the 1.88 expected events, 1.7 are predicted to come from the standard $\nu_\mu - \nu_\tau$ oscillations of the atmospheric sector, and 0.18 events come from various background sources simulating a $\nu_\tau$ interaction. Given the small numbers involved, I neglect the systematic error on the 1.88 expected events with respect to the statistical fluctuations, and compute the 90% Confidence Level (CL) upper limit (u.l.) on the signal, for the case of 3 observed events with 1.88 expected background. The signal is represented in this case by the number of additional $\nu_\tau$ CC interactions. Following ref.[4], I get un upper limit of 5.5 events.

To transform the 5.5 events u.l. into a probability, I use the equation:

$$A \times \int \phi(E) \times E \times R(E) \times P_{extra}(E) dE = 5.5$$

where A is a normalization factor discussed later; $\phi(E)$ is the neutrino flux at the OPERA detector, as given in ref.[4]; E is the neutrino energy, used to reproduce the behaviour of the $\nu_\mu$ CC cross section; R(E) is the ratio $\sigma_\tau(E)/\sigma_\mu(E)$ of the CC cross section for $\nu_\mu$ and $\nu_\tau$. $P_{extra}$ is the quantity to be determined, and represents the probability of conversion of $\nu_\mu$ into $\nu_\tau$. The normalization factor A is computed considering that in [2], the 1.7 events expected from the standard oscillation result from using a probability of the form: $P_{atm}=\sin^2(2\theta_{atm})\sin^2(1.27\Delta m^2_{atm}(L/E))$, with $\sin^2(2\theta_{atm})=1$. and $\Delta m^2_{atm}=2.32 \times 10^{-3}$ eV$^2$. Then, the normalization factor A is given by the equation:

$$A \times \int \phi(E) \times E \times P_{atm}(E) dE = 1.7$$

(I have assumed a constant A, which implies that the detection efficiency for $\nu_\tau$ interactions does not depend from the energy).

To compute the limit, I first considered the case of a probability $P_{extra}$ independent of E. In this case the result is $P_{extra} < 0.041$. In other words, assuming an energy independent probability of $\nu_\tau$ appearance, it is found that, besides the standard atmospheric oscillation, the flux of $\nu_\tau$ at the detector is not larger than 4.1% of the $\nu_\mu$ flux.

I have then investigated the limit that can be set when using an oscillation probability of the simple form: $P_{extra}=\sin^2(2\theta_X)\sin^2(1.27\Delta m^2_X(L/E))$. Doing so, one obtains the usual exclusion plot in the $\sin^2(2\theta_X)$-$\Delta m^2_X$ plane. The 90% CL limit resulting from this analysis is shown in indigo in Fig.1.

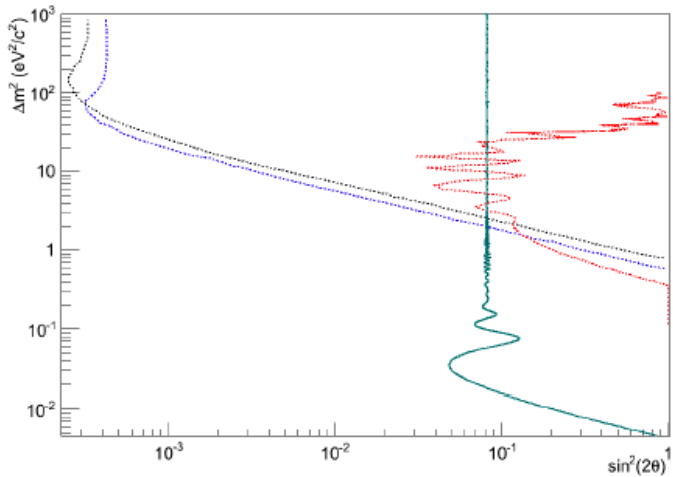

Figure 1: The indigo curve represents the 90% CL limit from this analysis. Also shown, are the limits from the short baseline $\nu_\tau$ appearance experiments NOMAD [5] (black curve) and CHORUS [6] (blue curve), and from the MiniBooNE-SciBooNE [7] $\nu_\mu$ disappearance experiment (red curve).

For large $\Delta m^2$, the 90% CL upper limit on $\sin^2(2\theta_X)$ is $8.2 \times 10^{-2}$. In Fig.1 are also shown the most stringent limits of direct searches for $\nu_\mu - \nu_\tau$ oscillations, obtained by the NOMAD [5] and CHORUS [6] experiments, with high sensitivity for $\Delta m^2$ values larger 10 eV$^2$. Indirect searches were also performed by studying $\nu_\mu$ disappearance, by experiments for which the oscillation would result in a reduction of the number of events with a muon in the final state. The most stringent limit has been obtained from the joint effort of the MiniBoone and SciBoone Collaborations [7] and is displayed in red in Fig. 1. It must be noted that the experiments mentioned above have a much smaller L/E ratio than OPERA, and are therefore, differently from OPERA, totally insensitive to the atmospheric oscillation. Then, for a direct comparison the results, one has to assume that in the case of OPERA the relatively small, but not negligible contribution from the atmospheric oscillation can be subtracted incoherently. Nevertheless, Fig.1 shows that the present analysis gives a result which is unique in the 0.005 to 0.300 eV$^2$ range for $\Delta m^2$, and complements the searches for $\nu_\mu$ disappearance for $\Delta m^2$ between 0.3 and 2.0 eV$^2$. The latter region is particularly interesting for some models involving sterile neutrinos [3], and relating to the LSND [8] and MiniBooNE [9] anomalies. Studies of different oscillation channels (including $\nu_\mu - \nu_\tau$) with models involving sterile neutrinos can be found in the literature; in particular, in [10] explicit reference is made to the configuration of the CNGS beam. Here however, I do not attempt to interpret the results in terms of any specific model.

It is finally useful to recall that, in a similar way to that described in this paper, ICARUS and OPERA have recently searched for an excess of $\nu_e$ interactions in the CNGS beam at the Gran Sasso [4,11,12]. Also in that case, no excess has been detected with respect to what expected from oscillations at the atmospheric $\Delta m^2$.